\documentclass[aps,prd,twocolumn,nofootinbib,10pt,floatfix,preprintnumbers]{revtex4-1}
\usepackage{amsmath}
\usepackage{amssymb}
\usepackage{verbatim}
\usepackage{orcidlink}
\usepackage[english]{babel}
\usepackage{xcolor}
\usepackage{graphicx}
\usepackage{dcolumn}
\usepackage{bm}
\usepackage[separate-uncertainty=true,multi-part-units=single]{siunitx}
\usepackage{numprint}
\usepackage{hyperref}
\hypersetup{
     colorlinks=true,
     breaklinks=true,
     linkcolor=blue,
     filecolor=blue,
     citecolor=blue,      
     urlcolor=cyan,
     }
\usepackage[caption=false]{subfig}

\newcommand{\Trh}{T_{\mathrm{RH}} }

\newcommand{\Neff}{N_{\mathrm{eff}} }

\DeclareSIUnit[quantity-product = {}]\parsec{\text{pc}}
\graphicspath{{./}{figures/}}

\begin{document}

\title{Visible in the laboratory and invisible in cosmology: decaying sterile neutrinos}

\preprint{UCI-HEP-TR-2023-09}

\author{Kevork N.\ Abazajian\,\orcidlink{https://orcid.org/0000-0001-9919-6362}}
\email{kevork@uci.edu}

\author{Helena Garc\'ia Escudero\,\orcidlink{https://orcid.org/0000-0003-4322-1607}}
\email{garciaeh@uci.edu}

\affiliation{Center for Cosmology, Department of Physics and Astronomy, University of California, Irvine,\\
Irvine, California 92697-4575, USA }
\date{\today}

\begin{abstract}
The expansion history and thermal physical process that happened in the early Universe before big bang nucleosynthesis (BBN) remains relatively unconstrained by observations. Low reheating temperature universes with normalcy temperatures of $T_\mathrm{RH}\sim 2\,\mathrm{MeV}$ remain consistent with primordial nucleosynthesis, and accommodate several new physics scenarios that would normally be constrained by high-temperature reheating models, including massive sterile neutrinos. We explore such scenarios' production of keV scale sterile neutrinos and their resulting constraints from cosmological observations. The parameter space for massive sterile neutrinos is much less constrained than in high-$T_\mathrm{RH}$ thermal histories, though several cosmological constraints remain. Such parameter space is the target of several current and upcoming laboratory experiments such as TRISTAN (KATRIN), HUNTER, MAGNETO-$\nu$, and PTOLEMY. Cosmological constraints remain stringent for stable keV-scale sterile neutrinos. However, we show that sterile neutrinos with a dark decay to radiation through a $Z^\prime$ or a new scalar are largely unconstrained by cosmology. In addition, this mechanism of sterile neutrinos with large mixing may provide a solution to the Hubble tension. We find that keV-scale sterile neutrinos are therefore one of the best probes of the untested pre-BBN era in the early Universe and could be seen in upcoming laboratory experiments.
\end{abstract}
\maketitle

\section{Introduction} \label{sec:intro}

Neutrino oscillations provide strong evidence for non-zero neutrino masses and are one of the most clear pieces of evidence of physics beyond the Standard Model (SM) \cite{deGouvea:2016qpx}. By measuring the fluxes and energy spectra of neutrinos coming from various sources, such as the Sun, nuclear reactors, and cosmic rays interacting with the Earth's atmosphere, numerous experiments have provided compelling evidence for neutrino oscillations and, by extension, non-zero neutrino masses \cite{Workman:2022ynf}.\\

In near unanimity, models for neutrino mass generation require the presence of new sterile neutrino states through either Majorana or Dirac neutrino mass mechanisms \cite{deGouvea:2016qpx}. Specifically, the addition of two sterile neutrinos can explain both solar and atmospheric neutrino oscillations, while a third massive sterile neutrino has considerable freedom as to its mass and mixing properties \cite{deGouvea:2005er,Asaka:2005an,Lindner:2010wr}, and can be a natural dark matter candidate \cite{Dodelson:1993je}. In this scenario, the sterile neutrino would be a neutral particle that does not participate in the Weak interactions, but could be produced by neutrino oscillations or other mechanisms, and could survive from the early Universe to the present day as a dark matter candidate \cite{Abazajian:2017tcc}. The mass scale of the sterile neutrino would need to be in the range of a few to tens of keV in order to be consistent with the observed properties of dark matter \cite{Zelko:2022tgf}.  
Sterile neutrinos can also affect neutrino oscillation experiments by introducing additional oscillation channels and modifying the observed oscillation patterns. Although these particles have not been definitively detected, on-going experiments have reported anomalies that could potentially be explained by these particles \cite{LSND:1996ubh,LSND:1997vqj,LSND:2001aii,MiniBooNE:2007uho,MiniBooNE:2010bsu, MiniBooNE:2010idf}. Further experimental investigations are ongoing to explore the possibility of sterile neutrinos and their role in neutrino physics \cite{Acero:2022wqg}.\\

The predominant model for the early Universe postulates that it underwent inflation, which diluted any prior constituents to cosmological irrelevance, assured cosmological flatness, and created the primordial density perturbations. When inflation comes to an end, its potential steepens, violating the slow roll, leading to the beginning of the reheating. During this phase, all particles that are kinematically permitted are directly created or generated through the thermal bath that the inflaton decay creates.  The reheating temperature, $\Trh$, refers to the temperature of the Universe after the period of inflation when particle decays transfer their energy into SM thermalized particles, establishing the initial hot and dense state.  Radiation domination evolution leading into the required era of big bang nucleosynthesis (BBN) places the lower limit of the cosmological reheating temperature to be as low as $\Trh = 1.8\,\mathrm{MeV}$ and be consistent with primordial nucleosynthesis, and with new physics that adds relativistic energy density, can be consistent with all observations \cite{Hannestad:2004px,Hasegawa:2019jsa}. There are a variety of histories prior to reheating (e.g., kination and scalar-tensor cosmologies) \cite{Rehagen:2014vna, Gelmini:2019esj,Allahverdi:2010xz}, and even the lowest $\Trh$ models can accommodate other key important components required of the early Universe, {\it viz.} baryogenesis and dark matter production \cite{Giudice:2000ex}.  

Cosmology, therefore, permits $\Trh$ to be anywhere from above the Grand Unified Theory scale $\Trh\gtrsim 10^{15}$ GeV and the weak freezeout or BBN scale of $\Trh \sim 2\,\mathrm{MeV}$, so that $\Trh$ remains a frontier for cosmology. In the case of a high scale, the long period of weak scattering at high $T$ allows for the thermalization of sterile neutrinos that oscillate with the active neutrinos for much of the parameter space of interest for neutrino oscillations at the eV to sub-eV sterile neutrino mass scale \cite{Langacker:1989sv}. For high $\Trh$, $\sin^2(2\theta)$ is tightly constrained, typically  $\sin^2(2\theta)< 10^{-7}$, to ensure that the production of sterile neutrinos in the early Universe is suppressed. However, for sufficiently low $\Trh$ universes, the scattering epoch is significantly reduced, so that even short-baseline-motivated eV-scale sterile neutrinos are allowed \cite{Hasegawa:2020ctq}. 

\begin{figure}[t!]
\centering
  \includegraphics[width=3.42in]{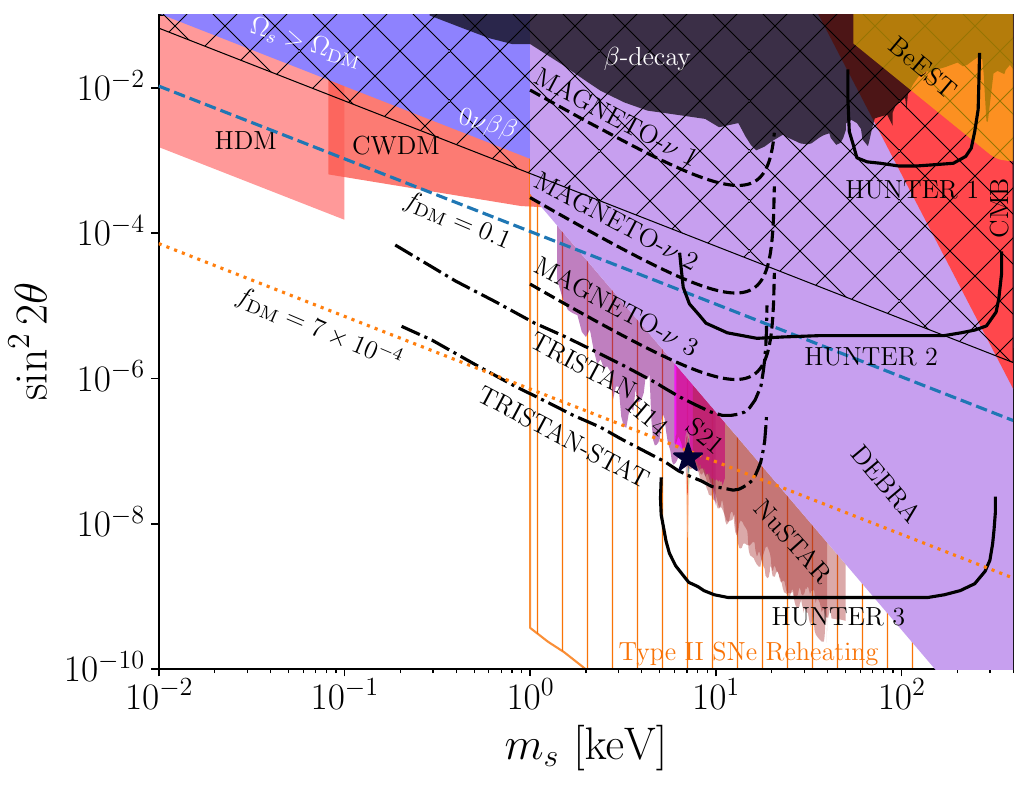}
  \caption{\footnotesize Shown here is the parameter space for a possible low reheating temperature universe with $\Trh = 5\,\mathrm{MeV}$, for the case of $\nu_s\leftrightarrow\nu_e$ mixing. The regions of this figure are described in the beginning of Sect.~\ref{sec:Current Constraints}. \label{fig:standardLRT} }
\end{figure}

In such low reheating temperature (LRT) universes, keV-scale sterile neutrinos with larger mixing angles are also allowed for much of their parameter space \cite{Gelmini:2004ah, Rehagen:2014vna, Gelmini:2019esj}. Importantly, sterile neutrinos at the keV-scale in LRT models cannot be the dark matter (see discussion in Sec.~\ref{sec:Current Constraints}). However, sterile neutrinos in LRT universes still undergo radiative decay, which can be detected by astronomical X-ray telescopes, as in the case of high reheating temperature (HRT) universes \cite{Abazajian:2001nj,Abazajian:2001vt}, and could even be responsible for the unidentified X-ray line at $\sim$3.5 keV \cite{Bulbul:2014sua,Boyarsky:2014jta}. In Ref.~\cite{Benso:2019jog}, the authors consider several mechanisms to decouple astrophysical and cosmological constraints on large-mixing angle keV-scale sterile neutrinos, including cancellation of the $\nu_s$ decay rate with new particles, new particles that mediate $\beta$-decay differently than $\nu_s$ decay, CPT violation, lepton number suppression, as well as suppression of production of sterile neutrinos in LRT universes with an additional reduction of their contribution to the dark matter density with no specified mechanism (that paper's ``cocktail'' model). However, if they are not associated with any other new beyond the Standard Model (BSM) physics, sterile neutrinos in LRT universes remain significantly constrained from radiative decay and structure formation, which we show in Fig.~\ref{fig:standardLRT}.  \\

Sterile neutrinos are a BSM extension that may be embedded in a richer phenomenology of their dark sector. It is known that cosmological constraints on active neutrino masses can be alleviated if the active neutrinos annihilate \cite{Beacom:2004yd} or decay \cite{FrancoAbellan:2021hdb}. Similarly, sterile neutrinos that are partially or fully thermalized in the early Universe may decay into lighter states through a new $Z^\prime$, $\nu_s\rightarrow \nu_{s^\prime}+\bar\nu_{s^\prime} +\nu_{s^\prime}$, or through a new scalar $\nu_s\rightarrow \nu_{s^\prime}+\phi$ \cite{Dolgov:2002wy}, altering their cosmological impact and related constraints. 

In this paper, we study the decay of keV-scale sterile neutrinos to a dark sector, which can allow for their presence at larger mixing angles. Interestingly, the disparate redshifting of sterile neutrinos when they are relativistic vs. non-relativistic can augment low cosmological relativistic energy density, $\Neff$, resulting from LRT models to match that inferred from cosmic microwave background (CMB) and large-scale structure observations. ($\Neff$ is defined in Sec.~\ref{sec:dark_model}.) In addition, this mechanism can provide a higher $\Neff$ to match that preferred by the Hubble tension (e.g., see \cite{Gelmini:2019deq,Escudero:2022rbq}). Further, we show how dark decay of the sterile neutrino, when combined with LRT, opens up all of the parameter space of interest for nuclear decay searches for keV-scale sterile neutrinos, including HUNTER \cite{Smith:2016vku,Martoff:2021vxp}, TRISTAN \cite{KATRIN:2018oow}, MAGNETO-$\nu$ \cite{xianyi_zhang_2022_6805550,MAGNETOnu}, and PTOLEMY \cite{PTOLEMY:2019hkd,Choi:2022gbs}. If a keV-scale sterile neutrino is detected in the parameter space in which these experiments are sensitive, it would be a new probe of the pre-BBN epoch and indicate new physics in the early Universe. In Sec.~\ref{sec:Current Constraints}, we briefly review LRT models and the constraints on keV-scale sterile neutrinos that mix with the active neutrinos. We introduce two dark decay models and show how dark decay can enhance the cosmological consistency of LRT models in Sec.~\ref{sec:dark_model} as well as provide a solution to the Hubble tension. We conclude in Sec.~\ref{sec:conclusions}.

\section{Neutrinos in a Low-Reheating Temperature Universe}
\label{sec:neutrinos_in_LRT}

The active neutrinos remain in contact with the plasma through weak interactions until temperatures of $\sim\!2$ to 5 MeV---the so-called temperature of weak decoupling. For a standard lepton-number symmetric background, sterile neutrinos are produced through oscillation-based scattering production at the highest rates at $T\approx 130\, \mathrm{MeV} (m_s/1\,\mathrm{keV})^{1/3}$ \cite{Dodelson:1993je}. However, sterile neutrinos can still be produced in the epoch of weak decoupling when their mixing is sufficiently large. Therefore, keV-scale sterile neutrinos are still subject to cosmological constraints at the largest mixing angles \cite{Gelmini:2004ah,Gelmini:2019esj}.  In LRT cases, the cosmological relativistic energy density, $N_\mathrm{eff}$, in active neutrinos is  reduced. Even though LRT universes of $\Trh = 1.8\, \mathrm{MeV}$ are consistent with BBN \cite{Hasegawa:2019jsa}, they produce $\Neff = 1.0$, which is highly discrepant with $\Neff$ determined from Planck's observations of the CMB and galaxy surveys' baryon acoustic oscillations, which find $N_\mathrm{eff}=2.99^{+0.34}_{-0.33}$ (95\% CL) \cite{Planck:2018vyg}. For reheating temperatures close to $\Trh= 5\ \mathrm{MeV}$, $N_\mathrm{eff}$ can be within 10\% of its canonical value, so that it remains consistent with current constraints from the CMB and large scale structure. LRT universes with $\Trh = 1.8\,\mathrm{MeV}$ are still possible when there is another source for relativistic energy density. In this and other scenarios we explore below, extra relativistic energy density contributing to $\Neff$ is from the decay of massive sterile neutrinos. 

For the case of LRT universes, the production of sterile neutrinos proceeds via partial thermalization due to low temperatures, which accommodates larger mixing angles than in HRT universes. Following the notation of Ref.~\cite{Gelmini:2004ah}, the $\nu_s$ distribution function produced in the early Universe turns out to be
\begin{equation}
f_s(E,T) \approx 3.2\,
d_\alpha\left(\frac{\Trh}{5\,\mathrm{MeV}}\right)^3\sin^2 2\theta
\left(\frac{E}{T}\right) f_\alpha(E,T)\,,
\end{equation}
where $\sin^2 2\theta$ is the mixing angle between active and sterile neutrino states, $d_\alpha = 1.13$ for $\nu_\alpha = \nu_e$, and $d_\alpha=0.79$ for $\nu_\alpha = \nu_{\mu,\tau}$. The fraction of the sterile neutrino distribution produced is then
\begin{equation}
  f\equiv\frac{n_{\nu_s}}{n_{\nu_\alpha}} \approx 10\, d_\alpha\sin^2 2\theta\left(\frac{\Trh}{5\,\mathrm{MeV}}\right)^3\,.
\label{eq:fractionsterile}
\end{equation}
This scattering-based non-resonant production mechanism for the sterile neutrinos is the minimal case we consider in this work, and represents the conservative level of sterile neutrinos in LRT universes.

\section{Current Constraints} 
\label{sec:Current Constraints}

In this section, we discuss constraints, regions of interest in the mass-mixing plane, and potential signals for sterile neutrinos in the case of a LRT universe. At very high mixing angles, $\sin^2 2\theta \sim 0.1$, BBN is affected due to thermalization of the sterile neutrinos and their contribution to the relativistic energy density through BBN \cite{Gelmini:2004ah,Gelmini:2019wfp}. These constraints are above (weaker) than the other constraints we consider. A fundamental cosmological constraint comes from the exclusion of $\Omega_s>\Omega_\mathrm{DM}$\footnote{Here, $\Omega_i\equiv \rho_i/\rho_\mathrm{crit}$, where $\rho_\mathrm{crit}$ is the critical density of the Universe} at large mixing angles that overproduces the sterile neutrinos to be above the dark matter density, and this is shown in the blue region in Fig.~\ref{fig:standardLRT}. The edge of this region represents where $\Omega_s=\Omega_\mathrm{DM}$. However, this line is excluded by hot dark matter (HDM) constraints \cite{Ade:2015xua}, up to masses at which the sterile neutrinos act as warm dark matter (WDM) ($m_s\sim 0.1\,\mathrm{keV}$). Above the 0.1 keV mass scale, pure WDM constraints exclude the possibility of sterile neutrinos as the totality of dark matter: WDM constraints on Dodelson-Widrow nonresonantly-produced sterile neutrino dark matter are at the level of $\gtrsim$80 keV, from combined lensing plus galaxy counts constraints \cite{Nadler:2021dft,Zelko:2022tgf}. Since LRT scattering-produced sterile neutrinos are kinematically more energetic (i.e., ``hotter'') \cite{Gelmini:2019wfp}, then constraints on LRT sterile neutrinos are more stringent than 80 keV, and well into the diffuse extragalactic background limit at 1 keV along the $\Omega_s=\Omega_\mathrm{DM}$ line. This exclusion is independent of $\Trh$ within LRT models ($\Trh \lesssim 7\,\mathrm{MeV}$). The combined HDM and WDM constraints therefore exclude LRT models from producing sterile neutrinos as all of the dark matter, when combined with the diffuse extragalactic background radiation constraints (discussed below) \cite{Gelmini:2019wfp}. When $\Omega_s<\Omega_\mathrm{DM}$, mixed cold plus warm dark matter (CWDM) constraints are relevant \cite{anderhalden:2012jc}. Sterile neutrinos as fractions of the dark matter are therefore constrained by HDM and CWDM considerations, and these are shown in Fig.~\ref{fig:standardLRT}. Note that sometimes the HDM constraint is extended to masses $m_s>0.1\,\mathrm{keV}$, up to even 10~keV \cite{Hasegawa:2020ctq}, but that is inaccurate as the sterile neutrinos are considered to be WDM above approximately $m_s \sim 0.1\,\mathrm{keV}$, and either WDM or mixed CWDM limits become appropriate.  

Below the edge where $\Omega_s=\Omega_\mathrm{DM}$, sterile neutrinos comprise a fraction of the dark matter such that $f_\mathrm{DM}\equiv \Omega_s/\Omega_\mathrm{DM}$. We show two representative cases of $f_\mathrm{DM} = 0.1$ and $f_\mathrm{DM} = 7\times 10^{-4}$ in Fig.~\ref{fig:standardLRT}. The lower fraction is commensurate with central values of the candidate signals of an X-ray line at approximately $3.55\,\mathrm{keV}$, seen in the Perseus galaxy cluster, stacked galaxy clusters \cite{Bulbul:2014sua}, and M31 \cite{Boyarsky:2014jta}.  

We calculate X-ray limits in the LRT parameter space using the fraction of dark matter as $\nu_s$ at each point in the parameter space. In Fig.~\ref{fig:standardLRT}, we show five X-ray constraints using the commensurate fractional dark matter in the parameter space: 
\begin{enumerate}
    \item An analysis of 51 Msec of \textit{Chandra X-ray Space Telescope} deep sky observations across the entirety of the sky, sensitive to the Milky Way halo signal, by Sicilian et al.~\cite{Sicilian:2020glg}, are shown in magenta and labeled S21;
    \item  M31 \textit{Chandra} observations analyzed by Horiuchi et al.~\cite{Horiuchi:2013noa} are shown in purple and labeled H14, which we adopt because of their wider energy range than the first \textit{Chandra} constriants;
    \item NuSTAR observations toward the Milky Way Galactic Bulge for higher masses \cite{Roach:2019ctw}, are shown in brown and labeled NuSTAR;
    \item NuSTAR observations of the full sky, sensitive to the Milky Way halo, are complementary to the prior NuSTAR constraints \cite{Roach:2022lgo}, and are also shown in brown and labeled NuSTAR; and,
    \item The conservative but broad-band constraints on excess electromagnetic diffuse emission is labeled as the diffuse extragalactic background radiation (DEBRA) limit \cite{Boyarsky:2005us}.
\end{enumerate}
We do not show constraints from Ref.~\cite{Dessert:2018qih} as the limits are a factor of $\sim$20 weaker than claimed, which was acknowledged within Ref.~\cite{Dessert:2018qih}, and in subsequent comments \cite{Abazajian:2020unr,Boyarsky:2020hqb}. And, we do not show limits from Ref.~\cite{Foster:2021ngm} as that work does not include instrumental and on-sky lines present at 3.3 and 3.7 keV in their stated limits. 
Another astrophysical consideration comes into play in the orange vertically hatched region, where sterile neutrinos deplete energy in the core of a Type II supernova \cite{Abazajian:2001nj,Hidaka:2006sg,Arguelles:2016uwb,Ray:2023gtu}, though portions of this region may also be responsible for supernova shock enhancement \cite{Hidaka:2006sg} or the origination of pulsar kicks \cite{Kusenko:2006rh}. 

We also show regions that are constrained by laboratory experiments, independent of any astrophysical or cosmological models, in Fig.~\ref{fig:standardLRT}. Constraints exist from neutrinoless double-beta decay searches in the hatched region labeled 0$\nu\beta\beta$ \cite{KamLAND-Zen:2016pfg}, though a cancellation may exist that alleviates this constraint \cite{Rodejohann:2000ne,Li:2011ss,Abada:2018qok}. We also show the constraints from a collection of nuclear beta decay kink searches in the solid black region labeled $\beta$-decay \cite{Olive:2016xmw}. Results from the $\beta$-decay search by BeEST are also shown in golden yellow \cite{Friedrich:2020nze}.

When photons are produced in the decay of sterile neutrinos before recombination, and when these photons are produced after the thermalization time $t_{\rm th} \simeq 10^6$ sec, then this can distort the thermal nature of the CMB spectrum \cite{ELLIS1992399,PhysRevLett.70.2661} (see e.g. the discussion in Ref.~\cite{Gelmini:2008fq}). The COBE FIRAS limit~\cite{1996ApJ...473..576F} on distortions of the thermal CMB rejects lifetimes $t_{\rm rec}> \tau > t_{\rm th}$, where $t_\mathrm{rec}$ is the recombination time. The red region in the upper right corner of Fig.~\ref{fig:standardLRT} shows the CMB distortion limits. 

Several current and upcoming laboratory experiments are sensitive to the parameter space of sterile neutrinos we are considering here. In Fig.~\ref{fig:standardLRT}, the black dot-dashed line is the forecast 1$\sigma$ sensitivity of time-of-flight measures from the TRISTAN detector on the KATRIN $\beta$-decay experiment, with the lower line showing their statistical limit \cite{KATRIN:2018oow}. The three dashed black lines show the sensitivity of the three stages of MAGNETO-$\nu$ \cite{xianyi_zhang_2022_6805550,MAGNETOnu} 
The solid lines are the forecast sensitivity for the upcoming K-capture experiment HUNTER (Heavy Unseen Neutrinos by Total Energy-Momentum Reconstruction), in its three stages \cite{Smith:2016vku,Martoff:2021vxp}. PTOLEMY is a tritium $\beta$-decay experiment aimed at detecting the cosmological relic neutrino background which is expected to start collecting data within few years, and may have sensitivity to this parameter space~\cite{PTOLEMY:2019hkd,Choi:2022gbs}. Ref.~\cite{Choi:2022gbs} provides event rates for this parameter space, but not sensitivity curve is available, so we do not show one for PTOLEMY. 

We presented constraints at $\Trh = 5\,\mathrm{MeV}$ in this section. Other values for $\Trh$ would change the constraint considerations to some degree. For lower $\Trh$, the constraint regions shift upward in $\sin^2 2\theta$, as less thermalization occurs at a given $\sin^2 2\theta$. Conversely for higher $\Trh$, as $\Trh$ approaches the peak of production of sterile neutrinos at $T\approx 130\, \mathrm{MeV} (m_s/1\,\mathrm{keV})^{1/3}$, the constraints of an HRT universe apply. As discussed in the introduction, LRT universes with $\Trh = 1.8\,\mathrm{MeV}$ are allowed when there is a new source for relativistic energy density, such as sterile neutrinos with a dark decay mode, which we now explore.

\section{Dark Decay Model} \label{sec:dark_model}

The population of partially to fully thermalized sterile neutrinos may not be cosmologically long-lived. In the cases of relatively large mixing that we consider, the sterile neutrinos may decay more rapidly into another sterile neutrino, $\nu_s^\prime$, plus other dark sector particles \cite{Dolgov:2002wy,Gariazzo:2014pja}. Such decays are known to alleviate constraints when they occur to the active neutrinos (e.g. \cite{FrancoAbellan:2021hdb}), and they will also alleviate constraints on sterile neutrinos. In one class of such models, a generic scalar, $\phi$, is introduced with an interaction Lagrangian associated with the decay of the keV-scale $\nu_s$ to an arbitrarily lighter $\nu_s^\prime$, $\nu_s \rightarrow \nu_s' \phi$:
\begin{equation}
\label{eq:scalar}
\mathcal{L} \supset \frac{g_{i,j}}{2}\bar{\nu_j}\nu_i\phi+ \frac{g'_{i,j}}{2}\bar{\nu_j}i\gamma_5\nu_i\phi+\text{h.c.}\,
\end{equation}
where $\nu_i$ and $\nu_j$ are the largely-sterile neutrino mass eigenstates, and $g^{(\prime)}_{i,j}$ are the scalar (pseudoscalar) couplings. Decays of keV-scale sterile neutrinos induced by this coupling are unconstrained except for the cosmological considerations we present below.

Another possible channel for sterile neutrino decay is $\nu_s \rightarrow \nu_s' \bar{\nu}_s' \nu_s'$, mediated by a new $Z^\prime$ boson:
\begin{equation}
\mathcal{L}^{\nu}_{Z^\prime} =g \sum_{\alpha}(\bar{\nu}_{\alpha,L}\gamma^{\mu}\nu_{\alpha, L}) Z_{\mu}^\prime\,, \label{eq:zprime}
\end{equation}
where $g$ is the coupling constant associated with the new SU(2) interaction, and $\alpha$ goes over the sterile neutrino states, which in our case is the minimal case of two.

Both of these models in Eq.~\eqref{eq:scalar} \& Eq.~\eqref{eq:zprime} introduce a new mechanism of sterile neutrino decay within a dark sector. For our interests in this work, only the lifetime of the decay associated with these new interactions, $\tau$, is important, as well as the requirement that the decay products are arbitrarily light, so as to act as dark radiation for all of cosmological history. The decay products of $\phi$, $\nu_s^\prime$ therefore act as pure dark radiation in contribution to $\Neff$, which we define generally as a combination of the radiation energy density in the active neutrinos $N_\mathrm{eff,act}$, plus the sterile neutrinos $N_\mathrm{eff,ster}$, plus any relativistic decay products of the sterile neutrinos, $N_\mathrm{eff,\ast}$,
\begin{equation}
    \Neff =  N_\mathrm{eff,act} + N_\mathrm{eff,ster} + N_\mathrm{eff,\ast}\, , \label{eq:neff} 
\end{equation}
where
\begin{align}
N_\mathrm{eff,act} \equiv& \frac{1}{\rho_\nu} \sum_i \frac{1}{4\pi^3}\int E(p) f_i(p) \mathrm{d}^3 p  \, ,\\ N_\mathrm{eff,ster} \equiv& \frac{1}{\rho_\nu}\sum_j\frac{1}{4\pi^3}\int E(p) f_j(p) \mathrm{d}^3 p \, . \label{eq:defneff}
\end{align}
Here, $\rho_\nu$ is the relativistic energy density in a thermal single neutrino species, $i$ sums over the partial or fully thermalized active neutrino species with energy distributions $f_i$, and $j$ sums over the energy densities of the relic stable $\nu_s^\prime$ and $\phi$. In general, only one of $N_\mathrm{eff,ster}$ and $N_\mathrm{eff,\ast}$ will be nonzero as the sterile neutrinos become nonrelativistic and then decay into dark radiation. In the case of the lowest LRT models, e.g. $\Trh \approx 1.8\,\mathrm{MeV}$, $N_\mathrm{eff,act} \approx 1$\cite{Hasegawa:2019jsa}, so that $\Neff$ is predominantly dark radiation, while in higher LRT models, e.g. $\Trh \approx 7\,\mathrm{MeV}$, $\Neff$ is predominantly active neutrinos ($N_\mathrm{eff,act}$), with dark sector particles ($N_\mathrm{eff,\ast}$) contributing a small perturbation. 

\subsection{Evolution of the Abundance of Decaying Sterile Neutrinos in an LRT universe}

In the case of sterile neutrinos with appreciably mixing, their abundance in an LRT universe is initially set by their approach to equilibrium after $\Trh$. This process is described by the Boltzmann equation. Their subsequent evolution is set by their redshifting as radiation and, when $T\lesssim m_s$, as matter components, followed by their subsequent decay. 

With no direct coupling to the reheating mechanism, sterile neutrinos are not present at $\Trh$, and are never in thermal equilibrium in the early Universe \cite{Yaguna:2007wi, Kusenko:2009up}. Nevertheless, there are different mechanisms by which the relic population of sterile neutrinos could have been produced subsequent to reheating \cite{Dodelson:1993je,Shi:1998km, Kusenko:2006rh,Petraki:2007gq}. In this paper, we focus on the minimal model in which the production of sterile neutrinos requires no new physics other than neutrino mass and mixing, and production arises from non-resonant flavor oscillations between the active neutrinos $\nu_\alpha$ of the SM and the sterile neutrino $\nu_s$, as originally proposed by Dodelson and Widrow (DW) \cite{Dodelson:1993je}.

In the LRT model, prior $\Trh$, the entropy in radiation and matter is not conserved and consequently, the $T$ dependence on the scale factor $a$ is different than the usual $T \propto a^{-1}$. In this scenario, prior to the radiation dominated standard epoch, a scalar field oscillates coherently around its true minimum and dominates the energy density of the Universe.  The decay of this scalar leads to nonthermal decay products that subsequently thermalize to a $T = \Trh$, followed by standard radiation domination and evolution (see e.g. Refs.~\cite{Kawasaki:1999na,Gelmini:2006pw,Gelmini:2006pq} ). 

Interactions of active neutrinos with the surrounding plasma during the oscillations act as measurements and force the propagating neutrino energy eigenstates into determinate flavor states, which with some probability results in a sterile neutrino. For the parameter space of interest here, the production rate is usually not fast enough for sterile neutrinos to thermally equilibrate, and the process is a freeze-in of the final abundance.

\begin{figure}[t!]
\centering
  \includegraphics[width=3.45in]{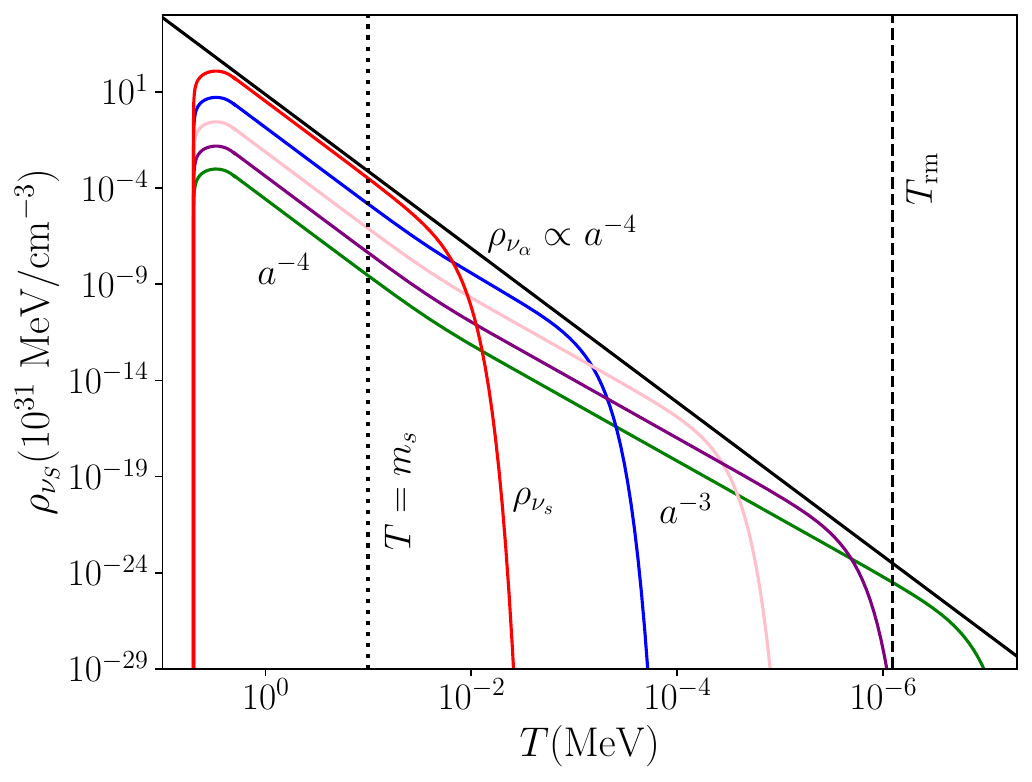}
  \caption{\footnotesize Shown here is the calculated production and energy density evolution for massless standard neutrinos (black line) and an example of $m_s= 100\,\mathrm{keV}$ sterile neutrinos. We show five different $\sin^2 2\theta$ cases of sterile neutrino energy density evolution, $\rho_{\nu_s}$. The sterile neutrinos decay at different times (temperatures) ranging from a case that matches the neutrino's mass $T_{\rm decay}= 0.1\,\mathrm{MeV}$ (red line) down to the temperature of matter radiation equality (green line). Note that sterile neutrinos with very different initial production densities can all match the same density relative to the active neutrinos at their decay. Therefore, a wide range of sterile neutrino masses and $\sin^2 2\theta$ can match a designated $\Neff$, when combined with either a nearly fully thermalized or non-thermalized $\rho_{\nu_\alpha}$ in LRT cosmologies. The reheating temperature in this example is chosen to be 5 MeV. The moments of $T=m_s$ and $T_\mathrm{rm}$ are shown with the dotted and dashed vertical lines, respectively. \label{fig:densityplot} }
\end{figure}

Assuming that only two neutrinos mix, $\nu_s$ and one active neutrino $\nu_\alpha$  ($\nu_e$ in all of the figures we present in this paper), the time evolution of the phase-space density distribution function of sterile neutrinos $f_{\nu_s}(p,t)$ with respect to the density function of active neutrinos $f_{\nu_{\alpha}}(p,t)$ is given by the following Boltzmann equation~\cite{Kolb:1990vq,Abazajian:2001nj}
\begin{align} 
\label{eq:boltzmann}
    \frac{d}{dt}f_{\nu_s}(p,t) ~=&~ \frac{\partial}{\partial t}f_{\nu_s}(p,t) - Hp\frac{\partial}{\partial p}f_{\nu_s}(p,t) \notag\\ ~=&~ \Gamma(p,t) \Big[f_{\nu_\alpha}(1-f_{\nu_s})-f_{\nu_s}(1-f_{\nu_\alpha})\Big]~.
\end{align}
Here $H$ is the expansion rate of the Universe, $p$ is the magnitude of the neutrino momentum and $\Gamma (p,t)$ is the conversion rate of active to sterile neutrinos. The active neutrinos are assumed to have a suppressed to full Fermi-Dirac distribution, depending on their thermalization state determined by $\Trh$.

Since $f_{\nu_s} \ll 1$ and $f_{\nu_s} \ll f_{\nu_{\alpha}}$, we take $(1-f_{\nu_s})=1$, and the second term in brackets on the right-hand side of Eq.~\eqref{eq:boltzmann} can be neglected. Thus, changing variables, Eq.~\eqref{eq:boltzmann} can be rewritten as ~\cite{Rehagen:2014vna,Abazajian:2001nj}
\begin{equation}
\label{eq:boltzmann2}
    -HT\left(\frac{\partial f_{\nu_s}(E,T)}{\partial T}\right)_{E/T} \simeq~ \Gamma(E,T)f_{\nu_\alpha}(E,T)~,
\end{equation}
where the derivative on the left-hand side is computed at constant $E/T$.

The conversion rate $\Gamma$ is the total interaction rate $\Gamma_{\alpha} = d_{\alpha} G_F^2 \epsilon T^5$ of the active neutrinos with the surrounding plasma weighted by the average active-sterile oscillation probability $\langle P_m \rangle$  in matter  (see Eq.~(6.5) of Ref.~\cite{Abazajian:2001nj}).

We obtain the sterile neutrino density distributions by integrating the Boltzmann Eq.~\eqref{eq:boltzmann2}. We show the resulting production of sterile neutrino density evolution at the far left of Fig. \ref{fig:densityplot}, where the density rises from zero. Here, we recover the results of Ref.~\cite{Gelmini:2019esj}.  Following production, the $\nu_s$ energy density dilutes as radiation, $\rho_{\nu_s} \propto a^{-4}$, as long as $T \gg m_s$. As the Universe cools to $T\sim m_s$, there is a transition of the decrease of the $\nu_s$ energy density to redshifting as matter, $\rho_{\nu_s}\propto a^{-3}$. For the example shown in Fig.~\ref{fig:densityplot}, $m_s = 100\,\mathrm{keV}$, but onset of pure matter-like redshifting occurs somewhat later than $T\approx m_s$, as LRT-produced sterile neutrinos are slightly ``hotter'' than thermal, with $\langle p\rangle \approx 4.11 T$ \cite{Gelmini:2019esj}.

The presence of sterile neutrinos with masses between the epoch of BBN and the photon last-scattering time allows the $\nu_s$ to augment their energy density with respect to the active neutrinos' by becoming nonrelativistic and redshifting more slowly. They then deposit their energy density back into relativistic dark decay products ($\nu_{s^\prime}$ and/or $\phi$) denominated as $N_\mathrm{eff,\ast}$. Therefore, the massive $\nu_s$ can boost $\Neff$ above $N_\mathrm{eff,act}$ produced by reheating alone in an LRT model, Eq.~\eqref{eq:neff}. The amount of relativistic energy deposited can be approximated by matching the density of the nonrelativistic sterile neutrinos with the targeted boost in dark radiation, 
\begin{equation}
    m_s n_{\nu_s}=N_\mathrm{eff,\ast}\rho_{\nu_{\alpha}}\, .
\end{equation}
Using Eq.~\ref{eq:fractionsterile}, we solve for the fraction of sterile neutrino production for a given $N_\mathrm{eff,\ast}$ as
\begin{equation}
   f=\frac{N_\mathrm{eff,\ast}\rho_{\nu_{\alpha}}}{m_s n_{\nu_{\alpha}}}\, . \label{eq:ftarget}
\end{equation}
This relation then stipulates what production of $\nu_s$ is needed to hit the target $N_\mathrm{eff,\ast}$. The energy density boost needed for smaller levels of production at smaller mixing angles requires longer matter-like redshifting before decay. We take the maximum decay time to be that corresponding to matter radiation equality, $T_\mathrm{rm}$, so that the decays do not directly affect the photon decoupling epoch. 

The relativistic energy density contribution of sterile neutrino decays can range from a majority of $\Neff$  ($N_\mathrm{eff,\ast} \gtrsim N_\mathrm{eff,act}$) to a small perturbation onto $\Neff$ ($N_\mathrm{eff,\ast} < N_\mathrm{eff,act}$), depending on the $\Trh$, $m_s$, $\sin^2 2\theta$, and $\tau$. Since $\Neff$ can be augmented by this mechanism, it may be responsible for any potential evidence for $\Neff$ above its standard value. For higher $\Trh$, more production occurs for a given $\sin^2 2\theta$ and $m_s$, so that higher $\Trh$ models probe smaller mixing angles (see Fig.~\ref{fig:DarkDecay_LRT}). We go through several examples in the following section. 

In summary, the sterile neutrinos can decay at a wide range of time scales, as shown in Fig. \ref{fig:densityplot}. As an example, we illustrate five decay timescale scenarios. The most rapid decay time is commensurate with $T = m_s$ (this case's outcome would be similar to any decay timescale at $T<m_s$, as the relativistic energy density in $\nu_s$ is simply transferred to the dark states). The slowest decay we consider occurs at the time of radiation-matter equality $T_\mathrm{rm}$. We also show three intermediate cases. 
In addition, we plot the energy density evolution of the massless active neutrinos (black line).

\subsection{Decaying Sterile Neutrino Parameter Space in LRT Cosmologies}

\begin{figure*}[t!]
\centering
\includegraphics[width=3.45in]{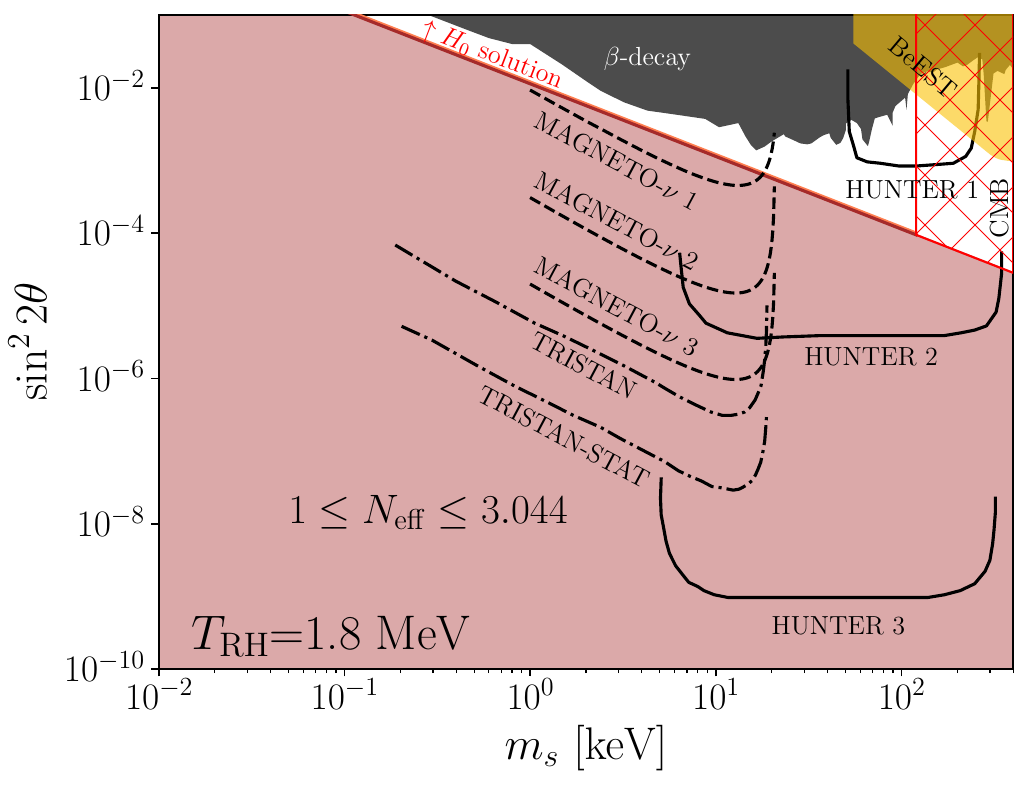}
  \includegraphics[width=3.45in]{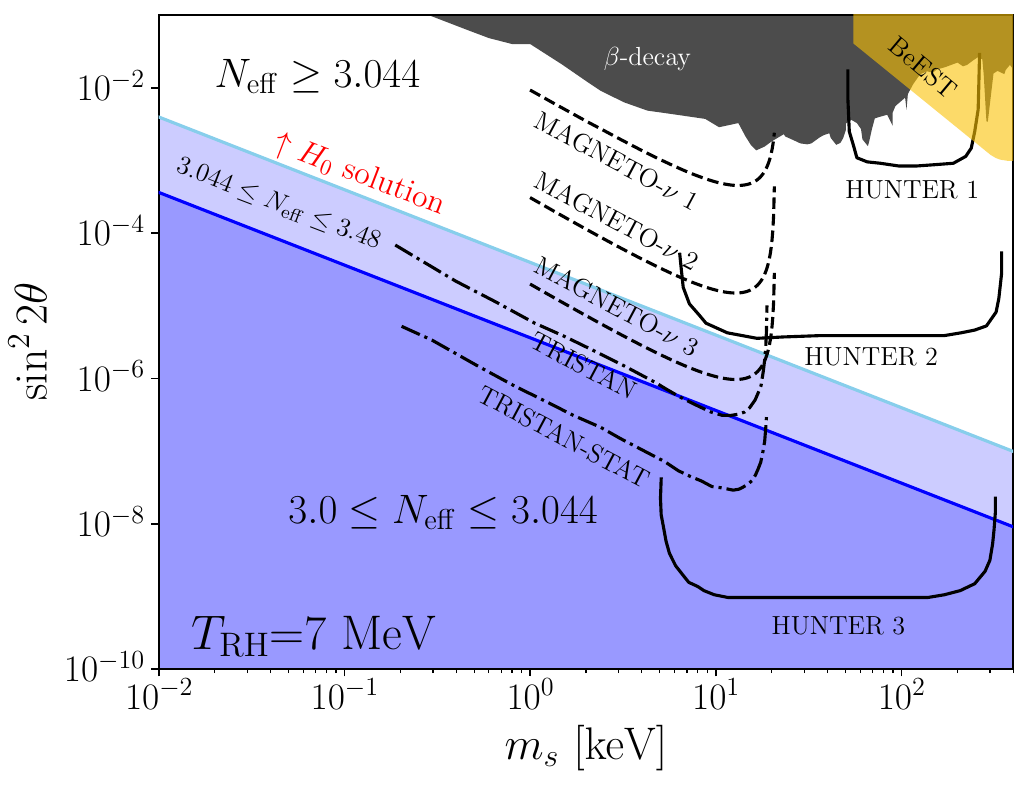}
  \caption{\footnotesize Shown are the updated parameter spaces of dark decay in an LRT Universe for two cases, $\Trh = 1.8\, \mathrm{MeV}\text{ and } 7\,\mathrm{MeV}$, left and right, respectively. The diagonal reddish and blueish lines correspond to the cases when the decay happens at the temperature of matter-radiation equality, and match different $\Neff$ values. For each pair, the darker color (lower) is associated with a value of $N_{\rm eff, Std} = 3.044$, and the lighter color one $N_{\rm eff, H0} = 3.48$ \cite{Escudero:2022rbq}. For the darker shaded region, $\Neff$ ranges from the minimal provided by $N_\mathrm{eff,act}$ to the standard value with a contribution from sterile neutrino decay. In the lighter shaded region on the right panel, $\Neff$ ranges from the standard value to that preferred to alleviate the Hubble tension, $N_\mathrm{eff, H0}$. Above the lighter diagonal, the  parameter space can  accommodate an alleviation of the Hubble tension with $N_\mathrm{eff, H0}$, or provide the standard density, $N_{\rm eff, Std}$.  For the case of $\Trh = 1.8\,\mathrm{MeV}$, the thermal nature of the CMB \cite{PhysRevLett.70.2661} constrains the red hatched portion. This constraint does not apply for $\Trh = 7\,\mathrm{MeV}$ in this parameter space.}
  \label{fig:DarkDecay_LRT}
\end{figure*}

In an LRT cosmology, sterile neutrinos are produced just after $\Trh$, then redshift as radiation, and potentially as matter, before ultimately decaying to dark radiation species. The amount of decay products' contribution to $\Neff$ varies with several parameters, as determined by Eq.~\eqref{eq:ftarget}. We consider three potential final values for $\Neff$ (Eq.~\eqref{eq:neff}):
\begin{itemize}
    \item $N_{\rm eff, Std} = 3.044$, the standard value with a standard enhancement from electron-positron annihilation  \cite{Bennett:2020zkv,Akita:2020szl};
    \item $N_{\rm eff, Upper}= 3.33$, the 95\% CL upper bound from Planck 2018 [Eq.~(67b)] in Ref.~\cite{Planck:2018vyg}];
    \item $N_\mathrm{eff, H0} = 3.48$, the central value preferred by solutions to the Hubble, $H_0$, tension \cite{Riess:2021jrx,Escudero:2022rbq}.
\end{itemize}
The precise value of  $N_{\rm eff,act}$ in LRT models depends on $\Trh$. We use the bottom panel of Fig.~1 in Ref.~\cite{Hasegawa:2019jsa}, for the relation between $\Trh$ and $N_{\rm eff,act}$.  For the highest temperature LRT cosmology we consider, $\Trh = 7\,\mathrm{MeV}$, the active neutrinos are almost fully thermalized, and $N_\mathrm{eff,act}  = 3.0$. Therefore, to match the standard $\Neff$, we require $N_{\rm eff,\ast}= 0.044$, while matching the $H_0$ tension requires $N_{\rm eff,\ast}= 0.48$, respectively. For a $\Trh = 1.8\, \mathrm{MeV}$, $N_{\rm eff, act}  = 1.0$, therefore, the standard density requires $N_{\rm eff,\ast}= 2.044$ and solving the $H_0$ tension requires $N_{\rm eff,\ast}= 2.48$, respectively.

For the case where there is no matter-dominated evolution of the sterile neutrino before it decays, there is no energy boost from differential redshifting of active and sterile neutrinos, and all density must come from oscillation production.  Using Eqs.~\eqref{eq:fractionsterile} \& \eqref{eq:ftarget}, this corresponds to a value of $\sin^2 2\theta > 0.1$, above the parameter space we consider in Fig.~\ref{fig:DarkDecay_LRT}. Another limiting case is when the decay occurs at radiation-matter equality $T_\mathrm{rm}$. We calculate $T_\mathrm{rm}$ decay contours and plot them in Fig.~\ref{fig:DarkDecay_LRT}, for the reheating temperatures of 1.8 MeV (left panel) and 7 MeV (right panel). We show two lines: one where the active neutrino and sterile neutrino decay products' density matches the standard $N_{\rm eff, Std} = 3.044$ (darker, lower line)\cite{Planck:2018vyg} and where the active neutrino and sterile neutrino decay products' density matches the $H_0$ tension alleviating value of $N_\mathrm{eff, H0} = 3.48$ (lighter, upper line)\cite{Escudero:2022rbq,Riess:2021jrx}. In the right panel's lighter shaded area, $\Neff$ spans from its standard value to the value favoring the mitigation of the Hubble tension, denoted as $N_\mathrm{eff, H0}$. Above the lighter diagonal line, the parameter space is capable of supporting either a resolution to the Hubble tension with $N_\mathrm{eff, H0}$, or maintaining the standard density,  $N_{\rm eff, Std}$. That is, sterile neutrinos with mass and mixing above the the lighter curves are consistent with cosmology at the specified $\Neff$ values. This is achieved because there is a cancellation between enhanced production at higher $\sin^2 2\theta$ and earlier decay providing less matter-redshift boost (see Fig.~\ref{fig:densityplot}). 

Because the sterile neutrinos mix with active neutrinos, they have a loop radiative decay to a lighter predominantly-active neutrino mass eigenstate and a photon \cite{Shrock:1974nd,Pal:1981rm}. Since the decays take place necessarily during the photon-coupled era, the effects of the decay photon is on distorting the thermal nature of the CMB photons. Hu \& Silk \cite{PhysRevLett.70.2661} calculated spectral distortions to the CMB radiation originated by the decay of unstable relic particles during the thermalization epoch. The appropriate constraints are from Fig. 1 in Ref.~\cite{PhysRevLett.70.2661}, where these limits constraints are most stringent for the largest masses in late-decay scenarios (maximizing the coefficient of the $y$-axis in Fig.~1 of Ref.~\cite{PhysRevLett.70.2661}). For the models plotted in Fig.~\ref{fig:DarkDecay_LRT}, the decay happens at $T_\mathrm{rm}$, or an age of the Universe of $t_\mathrm{rm}\approx 1.6\times 10^{12}\,\mathrm{sec}$.  The limits in Ref.~\cite{PhysRevLett.70.2661} are presented in $y \equiv m_X(b n_X/n_\gamma)$, where $m_X$ is the decaying particle mass, $n_X/n_\gamma$ is its abundance relative to the photons, and $b$ is the branching ratio to photons for the decay. For the highest mass of our parameter space, $m_s = 400\,\mathrm{keV}$, $y=4.6\times 10^{-15}\,\mathrm{GeV}$ for $\Trh = 7\,\mathrm{MeV}$. However, for $\Trh = 1.8\,\mathrm{MeV}$, the abundance of $\nu_s$ is much greater at late times, and $y=1.3\times 10^{-11}\,\mathrm{GeV}$. So, no radiative decay constraint exists for the case of $\Trh = 7\,\mathrm{MeV}$. We find the highest $m_s$ compatible with the $\Trh = 1.8\,\mathrm{MeV}$ curve, and that is 120 keV, and the constraint is independent of $\sin^2 2\theta$. We show the CMB thermal constraint in Fig.~\ref{fig:DarkDecay_LRT}, and it applies to $\Trh = 1.8\,\mathrm{MeV}$ cosmologies, and is non-existent in our parameter space for cases approaching $\Trh = 7\,\mathrm{MeV}$.

The only other constraints that are present in the parameter space in this scenario are the $\beta-$decay (black contour) \cite{Olive:2016xmw} and BeEST (yellow contour) \cite{Friedrich:2020nze}. As shown in  Fig.~\ref{fig:DarkDecay_LRT}, laboratory experiments such as HUNTER, TRISTAN, or MAGNETO-$\nu$ can detect the signal of sterile neutrinos in much of the allowed regions of this parameter space.

\section{Discussion \& Conclusions} \label{sec:conclusions}

The reheating temperature of the Universe is unknown beyond the requirement that it is $\Trh > 1.8\,\mathrm{MeV}$, as long as there is a new source of relativistic energy density in addition to the active neutrinos, and $\Trh \gtrsim 5\,\mathrm{MeV}$, in the case of no new physics \cite{Hasegawa:2019jsa}. Therefore, $\Trh$ is a free parameter in studies of the energy content arising from the hot big bang. Baryogenesis and dark matter production can be accommodated in the reheating process \cite{Giudice:2000ex,Chowdhury:2023jft}. In LRT cosmologies, the weak-coupling epoch is significantly reduced, suppressing active-sterile neutrino oscillations. As a result, regions of keV-scale sterile neutrinos' parameter space that were previously forbidden by cosmological or astrophysical constraints can become viable \cite{Gelmini:2004ah}. If the dark sector in which the sterile neutrino participates includes dark decay channels, we have shown here that the parameter space in LRT cosmologies is even more significantly alleviated. The energy density described by $\Neff$ in such dark-decay LRT cosmologies can have both pure, decay-produced, non-thermal, radiation components, as well as massive active neutrino components. High sensitivity to $\Neff$ will be provided by current and upcoming CMB experiments such as CMB-S4 \cite{Abazajian:2019eic}. The discovery of a sterile neutrino with parameters in this region could indicate a rich LRT thermal history. 

Several current and upcoming laboratory experiments are sensitive to keV-scale sterile neutrino parameter space, including 
HUNTER \cite{Smith:2016vku,Martoff:2021vxp}, TRISTAN \cite{KATRIN:2018oow}, MAGNETO-$\nu$ \cite{xianyi_zhang_2022_6805550,MAGNETOnu}, 
and PTOLEMY~\cite{PTOLEMY:2019hkd,Choi:2022gbs}. Laboratory direct dark matter detection experiments employing xenon, including LZ \cite{LZ:2022lsv} and  XENONnT \cite{XENON:2023cxc}, could be sensitive to our considered parameter space \cite{Campos:2016gjh}. However, only the case of all of the dark matter being sterile neutrinos has been considered, and the constraints proportionately alleviate for fractional dark matter models, with all cases lying above the DEBRA constraints in Fig.~\ref{fig:standardLRT}, but they could considerably improve.
The appreciable mixing between active and sterile neutrinos we consider here may also arise from non-standard interaction (NSI) searches. Currently, the constraints in this NSI parameter space are largely the ones we have shown: $\beta$-decay and $0\nu\beta\beta$-decay \cite{Berryman:2022hds}.

The next few years could provide the potential discovery of laboratory-accessible sterile neutrinos, whose existence is in conflict with HRT cosmologies, with the aforementioned $\beta$-decay, K-capture, and neutrino capture experiments. In this work, we showed that the presence of decaying keV-scale sterile neutrinos could also be indicated by the $H_0$ tension. The discovery of keV-scale sterile neutrinos with appreciable mixing would be an important finding for particle physics, astrophysics, and cosmology, not only for its own discovery, but it would alter the usual assumptions of the early Universe and provide a new paradigm.

\section{Acknowledgements}
We would especially like to thank Graciela Gelmini for detailed discussions, as  well as James Alvey, Z.~Chacko,  Philip Lu, Alex Kusenko, and Tim Tait for helpful discussions. We also thank the referee for helpful comments. KNA is partially supported by U.S. National Science Foundation (NSF) Theoretical Physics Program, Grants PHY-1915005 and PHY-2210283. HGE was supported in part by the UC Southern California Hub, with funding from the UC National Laboratories division of the University of California Office of the President. HGE  was partially supported by a fellowship from the “La Caixa” Foundation (ID 100010434). The fellowship code is LCF / BQ / AA19 / 11720045.


\twocolumngrid
\bibliography{bibl}{}

\end{document}